\documentclass[compsoc, conference, a4paper, 10pt, times]{IEEEtran}
\usepackage{balance}

\usepackage{cite}
\usepackage{amsmath,amssymb,amsfonts}
\usepackage{algorithmic}
\usepackage{graphicx}
\usepackage{textcomp}
\usepackage{xcolor}
\usepackage{booktabs}
\usepackage[hidelinks]{hyperref}
\usepackage[T1]{fontenc}
\usepackage[acronym]{glossaries}
\usepackage{pgfplots}
\usepackage{subcaption}
\usepackage[textsize=footnotesize, french]{todonotes}
\usepackage{listings}
\usepackage{lstlangarm}

\makeatletter
\newcommand\currentStyle@lstparam{}
\lst@AddToHook{Output}{\global\let\currentStyle@lstparam\lst@thestyle}
\lst@AddToHook{OutputOther}{\global\let\currentStyle@lstparam\lst@thestyle}
\makeatother

\makeatletter
\newcommand{\highlightcode}[2]{\currentStyle@lstparam \textcolor{#1}{#2}}
\makeatother

\usepackage{multirow}
\usepackage[nolist]{acronym}
\usepackage[english]{babel}
\usetikzlibrary{
  arrows,
  automata,
  backgrounds,
  calc,                     
  chains,
  decorations.pathmorphing, 
  decorations.pathreplacing,
  fit,                      
  math,
  matrix,
  mindmap,
  patterns.meta,
  patterns,
  positioning,
  scopes,
  shapes.gates.logic.US,
  shapes.geometric,
  shapes.geometric,
  shapes.symbols,
  shadows,
  spy,
  decorations.pathmorphing, 
  trees,
}

\newcommand{\shorturl}[1]{\texttt{\href{https://#1}{#1}}}

\begin{document}

\title{Faulting original McEliece's implementations is possible\\
  {\large How to mitigate this risk?}}

\author{\IEEEauthorblockN{Vincent Giraud}
\IEEEauthorblockA{\textit{Ingenico}\\
  \'Ecole Normale Sup\'erieure, DIENS, CNRS,\\
  PSL University, Paris, France\\
  Email: vincent.giraud@ens.fr}
\and
\IEEEauthorblockN{Guillaume Bouffard}
\IEEEauthorblockA{\textit{National Cybersecurity Agency of France (ANSSI),} \\
  \'Ecole Normale Sup\'erieure, DIENS, CNRS,\\
  PSL University, Paris, France\\
  Email: guillaume.bouffard@ens.fr}
}

\maketitle

\begin{abstract}

  Private and public actors increasingly encounter use cases where they need to implement
  sensitive operations on mass-market peripherals for which they have little or
  no control. They are sometimes inclined
  to attempt this without using hardware-assisted equipment,
  such as secure elements. In this case, the white-box attack model is
  particularly relevant and includes access to every asset, retro-engineering,
  and binary instrumentation by attackers. At the same time, quantum attacks are
  becoming more and more of a threat and challenge traditional asymmetrical
  ciphers, which are treasured by private and public actors.

  The McEliece cryptosystem is a code-based public key algorithm introduced in
  1978 that is not subject to well-known quantum attacks and that could be
  implemented in an uncontrolled environment. During the NIST post-quantum
  cryptography standardization process~\cite{nistpq}, a derived candidate
  commonly refer to as \emph{classic} McEliece was selected. This algorithm is
  however vulnerable to some fault injection attacks
  while \emph{a priori}, this does not apply to the original McEliece. In this
  article, we thus focus on the original McEliece cryptosystem and we study its
  resilience against fault injection attacks on an ARM reference
  implementation~\cite{SPA_McEliece}. We disclose the first fault injection
  based attack and we discuss on how to modify the original McEliece
  cryptosystem to make it resilient to fault injection attacks.

\end{abstract}


\acresetall{}

\section{Introduction}
For many sectors,
the implementation of sensitive operations, such as
authentication, payment, or protection of intellectual property, increasingly
targets personal embedded devices such as smartphones. These kinds of
peripherals are often called \ac{COTS} devices, since they aim for the general
public, and are designed with mass-market constraints such as a strong sale
price limitation. Hence, despite the potentially high number of executable
sources present on the platform, they often do not have hardware security
mechanisms such as secure elements or secure enclaves. Even when these features
are embedded in the system, they may not be accessible to third-party industrial
actors, for commercial reasons. Such stakeholders might thus try to develop
alternative ways to address the lack of trust in these environment, which is
also relevant with the rapid growth of \ac{BYOD} policies allowing employees to
use their own personal devices, such as smartphones, laptops, and tablets, for
work-related activities.

Providing sensitive operations mainly implies exploitation of
cryptographic assets.
They can be used to encrypt, decrypt, or sign
a content. A company choosing not to rely on hardware security
mechanisms must endorse full responsibility for the protection of
these mechanisms. This situation led to the definition of the
\emph{white-box attack model}, which refers to attackers having all powers
on the execution environment.
In essence, it refers to the case of trying to protect assets
on a platform where a hostile actor is root,
with the ability to
read, write, or instrument everything~\cite{conf/ches/BosHMT16}. In
particular, we refer to the implementation of cryptographic algorithms
in this context as the \ac{WBC} security
model~\cite{conf/ccs/ChowEJO02}. Companies operating these algorithms
face a significant challenge: preventing attackers from recovering
secrets in this type of scenario. In the case of the AES,
contests~\cite{whibox} have shown that implementations that are not
broken in less than two weeks in the \ac{WBC} attack model are very
rare. To increase this time period, protections will be embedded
against reverse engineering and binary instrumentation; however, these
measures tend to significantly increase the resulting binary size,
decrease performance, and thus, impact usability. In concrete use
cases, developers will put in place a complete replacement of the
sensitive code on a regular basis~\cite{conf/sstic/Thomas22}; however,
these updates are not easy to maintain owing to mobile network
limitations, mainly availability, cost and bandwidth. Another problem
affecting implementation of cryptography in the white-box model is
code porting, meaning efforts to copy them on another device to
circumvent protections.

In parallel in recent years, the risks associated to cryptographic attacks by
quantum computers received an increasing attention with the release of the first
quantum hardware. This mode of functioning disrupts the conventional premises of
computations and the axioms of cryptography. In this context, called
post-quantum cryptography, traditional asymmetrical ciphers are affected,
especially RSA, while the impact on symmetrical ciphers is rather
limited. In response, the \ac{NIST} initiated a process to determine public key
algorithms suitable for this model~\cite{nistpq}. Both French~\cite{anssipq} and
German~\cite{bsipq} information security agencies have introduced the need to
transition from the current cryptosystems to post-quantum-compatible ones. In
light of these elements, systems and services developed today should consider
this threat.

At this point, industrials wishing to implement asymmetrical cryptographic
algorithms in an uncontrolled environment or in \ac{COTS} face many
challenges: resisting as long as possible against attackers able to fully
instrument their implementation, resisting against quantum attacks, producing
sufficiently lightweight binaries, and updating them regularly through mobile
networks. 

In this study, we explored the security of the McEliece cryptosystem which is an
interesting candidate that is quantum-safe and seems be more suitable on
uncontrolled environments. In our study, we focused on the possibility of using
McEliece and the security problems of using this cryptosystem in the white-box
context. For the rest of this article, McEliece cryptosystem refers to the original
McEliece.

\subsection{The McEliece Cryptosystem}
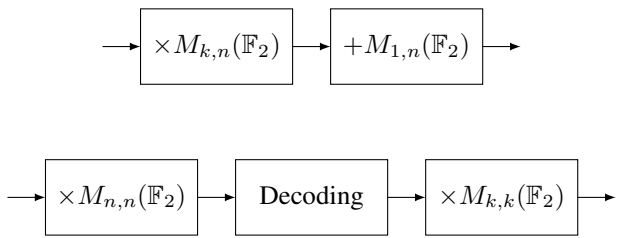
\begin{figure}[hbt]
\begin{center}
\begin{tikzpicture}
\def\hauteurB{1}
\def\largeurB{2}
\def\largeurF{0.5}
\def\espaceV{2}
\def\decalageH{(\largeurB + \largeurF)/2}

\draw ({\largeurF + \decalageH}, \espaceV) rectangle ++(\largeurB,\hauteurB) node[pos=0.5] {$\times M_{k,n}(\mathbb{F}_2)$};
\draw ({\largeurB * 1 + \largeurF * 2 + \decalageH}, \espaceV) rectangle ++(\largeurB,\hauteurB) node[pos=0.5] {$+ M_{1,n}(\mathbb{F}_2)$};
\draw [-latex] ({\decalageH },{\hauteurB * 0.5 + \espaceV}) -- ++(\largeurF,0);
\draw [-latex] ({\largeurB * 1 + \largeurF * 1 + \decalageH},{\hauteurB * 0.5 + \espaceV}) -- ++(\largeurF,0);
\draw [-latex] ({\largeurB * 2 + \largeurF * 2 + \decalageH},{\hauteurB * 0.5 + \espaceV}) -- ++(\largeurF,0);

\draw ({\largeurB * 0 + \largeurF * 1},0) rectangle ++(\largeurB,\hauteurB) node[pos=0.5] {$\times M_{n,n}(\mathbb{F}_2)$};
\draw ({\largeurB * 1 + \largeurF * 2},0) rectangle ++(\largeurB,\hauteurB) node[pos=0.5] {Decoding};
\draw ({\largeurB * 2 + \largeurF * 3},0) rectangle ++(\largeurB,\hauteurB) node[pos=0.5] {$\times M_{k,k}(\mathbb{F}_2)$};
\draw [-latex] (0,\hauteurB * 0.5) -- ++(\largeurF,0);
\draw [-latex] ({\largeurB * 1 + \largeurF * 1},\hauteurB * 0.5) -- ++(\largeurF,0);
\draw [-latex] ({\largeurB * 2 + \largeurF * 2},\hauteurB * 0.5) -- ++(\largeurF,0);
\draw [-latex] ({\largeurB * 3 + \largeurF * 3},\hauteurB * 0.5) -- ++(\largeurF,0);
\end{tikzpicture}
\end{center}
\caption{\small
  Representation of the operational principles in the McEliece cryptosystem.
  Above is the encryption, below is the decryption.
}
\label{fig:McEliece_internals}
\end{figure}

Introduced in 1978~\cite{mceliece_spec}, the McEliece cryptosystem is an
asymmetric cipher that employs a ($n$,$k$)-linear error-correcting code
$\mathcal{C}$ (correcting up to $t$ errors) with a fast decoding algorithm and
linear operations, as illustrated in Figure~\ref{fig:McEliece_internals}, where
$M_{a,b}(\mathbb{F}_2)$ denotes a matrix of size $a,b$, containing elements in
the finite field of order 2.  In this era, its large key sizes have made it less
favored than RSA; however, this criterion is now less significant because of
hardware developments, and we can now observe implementations tailored for
micro-controllers~\cite{microeliece}.

To encrypt a clear-text block $m$, it must be considered as a vector of $k$ bits, and
multiply it with a (public) matrix $G'=S\times G\times P$ of size $k,n$, which is the multiplication between
three distinct ones:
a random $k\times k$ invertible matrix $S$ used to
scramble the data;
$G$, a generator matrix for $\mathcal{C}$;
and a random $n\times n$ permutation matrix $P$ (\emph{i.e.} having a single
nonzero entry in each row and column).
Errors are then added voluntarily with a random vector $e$ of Hamming weight
below $t$ to compute the encrypted block $c = mG' + e$.
To decrypt
a cipher-text block $c$, one must be in possession of the three secret
matrices $S$, $G$ and $P$ separately,
to first cancel the permutation,
\emph{i.e.} compute $a = cP^{-1}$,
(quickly) decode $b$ from $a$ with the error-correcting code $\mathcal{C}$, and then
unscramble the data \emph{i.e.}, compute $m = bS^{-1}$.
An attacker intercepting the message $c$ would have to find the nearest codeword
of the code generated by $G'$ seen as a general linear code, which is known to be
NP-hard~\cite{linearcode_NBhard_78}.
An alternative offensive approach consists in attempting to recover the
structure of the underlying code $\mathcal{C}$. The original proposal relies on
binary Goppa codes, which so far remain one of the few families of codes which have largely
resisted attempts at devising such structural attacks.

In all cases, since the McEliece cryptosystem mainly consists of linear applications that
can be modeled as matrix multiplications, it has significant performance
potential. These operations can be accelerated by dedicated hardware, such a
\acp{GPU}, or by suitable instruction set extensions, particularly \ac{SIMD}
ones. These features are widely popular in different research topics and can be
applied directly in this context. However, many devices targeted by McEliece
implementations do not provide them, especially inexpensive micro-controllers.

The asymmetry of this algorithm is valued in the industry: having distinct
public and private keys provides more flexibility when designing use cases, and
is useful when conceiving secure communications or content delivery services,
for example.  This attribute has been presented by many other ciphers based on
factorization or lattices. However, these paradigms create many difficulties
when considering white-box implementations. To the best of our knowledge, this
mode of execution usually relies heavily on precomputed intermediate variables;
however, these are too large for the aforementioned algorithms. For example,
facto\-ri\-za\-tion implies operation on very large numbers.  For both
computation time and storage size reasons, precomputing them is completely out
of reach, making translation to the white-box model complex. The McEliece
cryptosystem, which embeds linear operations, provides opportunities to address
this issue.

\subsection{State-of-the-art McEliece Cryptosystem Security}

When exploring ciphers deemed reliable in a post-quantum context, the McEliece
cryptosystem is a frequent candidate. In the fourth round of the aforementioned
\ac{NIST} contest, which began at the end of 2016~\cite{nistpq}, one candidate
that still remained, \textit{Classic McEliece}, was inspired by the McEliece
specification. In the meantime, PQCRYPTO, a research group specialized in
cryptography in the post-quantum context, recommended the use of the original
McEliece cryptosystem, with only bigger keys than usual to consider the threat
of quantum-based attacks~\cite[section 4]{pqcrypto}. Additionally, the \ac{BSI}
also mentions it in its recommendations concerning quantum-safe
cryptography~\cite[page 29]{bsi_mceliece}.

In addition to its estimated robustness in a post-quantum context, this
cryptosystem provides good resistance against attacks by leveraging fault
injection. Indeed, when exploiting the private key during the decryption
process, a major operation is decoding the intermediary data using a linear
code, as illustrated in Figure~\ref{fig:McEliece_internals}. If one naively
applies corruption during the computation, the errors might be inherently
corrected, thus voiding the intrusion. On the one hand, this strength of the
McEliece cryptosystem is reflected in the scientific
literature~\cite{mceliece_faults} where authors focused on the theoretical
resistance of the algorithm against fault injection attack. They did not
succeeded in obtaining the private key. To the best of your knowledge, this work
is the only one where the original McEliece cryptosystem algorithm is studied against
hardware fault attacks.

On the other hand, \ac{NIST} candidate Classic McEliece, has been
the subject of such
attacks~\cite{cayrel_HDR}\cite{cryptoeprint:2022/1529}\cite{cryptoeprint:2021/840}.
The Goppa codes of a Niederreiter cryptosystem, which are derived from McEliece,
have been the subject of a fault injection based attack explained
in~\cite{Danner_2020}. These elements make it opportune to follow the PQCRYPTO
recommendations and focus on the original McEliece cryptosystem.

\subsection{The threat of hardware attacks on the \acl{WBC} model}
\label{sec:hard_wb}

The white-box security model assumes that attackers possess complete control
over an implementation, requiring various features to impede secret extraction
and make it as challenging and costly as possible. Binary
obfuscation~\cite{conf/uss/SchlogelBCABH022} is a feature that significantly
complicates reverse engineering. In the white-box model, robustness is essential
to enable implementations to maintain their reliability over an extended period,
resulting in fewer replacements and lower usage of mobile networks.

Hardware attacks form the basis of many effective offensives against white-boxes
implementations~\cite{conf/ches/BosHMT16}. When these attacks are translated
into the software world, they can often bypass the complexity and obfuscation
mentioned earlier, thereby jeopardizing the overall feasibility of white-box
implementations. As a result, what may have taken days or weeks to break can be
achieved in less than one day. \ac{DFA} on \ac{AES} serves as an example and is
applicable to both hardware~\cite{DFA_AES} and software~\cite{DFA_WB_AES}. From
perspective of an attacker, while the former provides an offensive against
an uncontrolled platform, the latter offers a significant operational shortcut
despite the controlled platform.

Research about execution perturbation is essential to reinforce today's
implementation prototypes. As part of this effort, offensive investigations
have been conducted on the McEliece cryptosystem because preventing fault
attacks is crucial. To simulate fault injection attacks easily, binary
instrumentation tools such as Rainbow\footnote{See
\shorturl{github.com/Ledger-Donjon/rainbow}} and QBDI\footnote{See
\shorturl{github.com/QBDI/QBDI}} can be employed. These tools enable the
efficient scrutiny of specific behaviors and precise modifications at specific
times. Moreover, these features can be leveraged to satisfy user-specified
real-time conditions.

\subsection{Contribution}

\noindent
Our main contributions are divided into three parts:

\begin{enumerate}
  \item we present a new fault injection based attack against a typical, common
        implementation~\cite{SPA_McEliece} of the McEliece cryptosystem on ARM
        targets;
  \item we discuss the applicability of this attack;
  \item we propose a variant of the McEliece cryptosystem expected to
    be intrinsically resistant to our attack and thus more suitable for
    use in uncontrolled environments as expected in the \ac{WBC} model.\newline
\end{enumerate}

The remainder of this article is organized as follows. In
Section~\ref{sec:mceliece-vs-fault}, we present an attack based on fault
injection that targets ARM reference implementation~\cite{SPA_McEliece} of the
McEliece cryptosystem. Section~\ref{sec:discussion} discusses how our
findings impact McEliece implementations in the \ac{WBC} security
model. In Section~\ref{sec:wbc-pq}, we introduce a variant of McEliece
cryptosystem to be immuned to our attack. Finally, Section~\ref{sec:conclusion}
offers concluding remarks and describes our future works.

\section{Faulting McEliece implementation}
\label{sec:mceliece-vs-fault}

We conducted an investigation into the vulnerability of the McEliece
cryptosystem for fault injection. As mentioned previously, this cipher is known
for its robustness against such efforts, and this aspect has already been
studied. However, because successful attempts are aimed at the \ac{NIST}
candidate named Classic McEliece, or towards the Niederreiter cryptosystem, we
decided, to push further the research into the original McEliece cryptosystem, to target
implementations. This initiative was linked to the study of the
possibility to implement it in uncontrolled environment. As explained in
\autoref{sec:hard_wb}, many successful attacks on these implementations result
from identical attacks intended for application on hardware. Identifying a
vulnerability through a fault injection attack will require protecting software
implementation against exploitation of this attack.

\subsection{Obstacles against efficient fault injections attacks}

In~\cite{SPA_McEliece}, Petrvalsky \textit{et al.} demonstrated that a
side channel attack can be conducted during the decryption process when matrix
multiplications are implemented using the conventional method. In the power
traces, one can identify the pattern corresponding to modulo 2 addition. When a
vector with a Hamming weight of 1 is inputted, it is possible to infer the position
of the only set bit of the corresponding line in the permutation matrix owing to
the temporal position of this pattern. Once the permutation matrix is recovered,
the rest of the private key can be mathematically computed.

We studied how matrix multiplication in the McEliece cryptosystem
reacts to fault injections. The main difficulty for an attacker is
that the recovery of information on intermediate data seems
impossible, as there are only two possibilities that can be observed
at the global output of the decryption process: either (1) the
decoding operation corrects the errors, making it appear as if the
fault never happened, or (2) we have modified the data too much,
resulting in an invalid or corrupted output. To overcome this
difficulty, we developed an offensive process that allows the
acquisition of information based on the final output of the algorithm,
either a correct or an error message.

The first step of the decryption process involves applying a permutation matrix
that has only one bit set in each row or column. The attacker's goal is to
determine the matrix. A fundamental property of these linear applications is
that they do not alter the Hamming weight, which means that the output will have
the same number of set bits as the input. Furthermore, we know the size of this
matrix, which is the ciphered block, denoted by $n$. Another important
parameter of a specific implementation of the McEliece cryptosystem is $t$,
the maximum number of bit errors the code can correct.

If we were able to scrutinize the output of this matrix, its recovery would be
easy. It would be sufficient to use an input vectors with only one-bit set and to
observe where they end up in the output. However, this is not possible because
the internals of the algorithm are either impossible to reach or intentionally
complex. Therefore, we investigated me\-thods to infer information about this part
based on the state of the global output.
However, this possibility seems mathematically out of
question, because the McEliece cryptosystem specifies the use of a scrambling
matrix to provide diffusion in the data, along with a decoding operation, which
is a surjective function.

\subsection{Attack methodology}

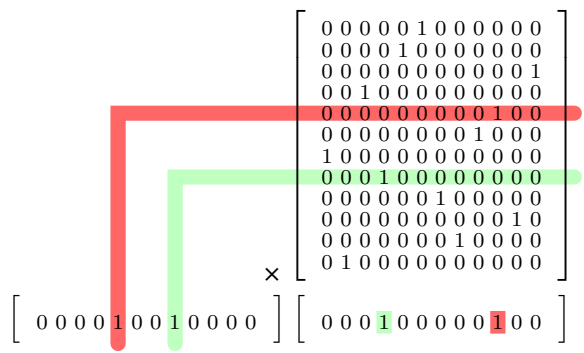
\begin{figure}[hbt]
\begin{center}
\begin{tikzpicture}
\def\Xdroit{3.75}
\def\Ydroit{2.35}

\matrix (m) [left delimiter=[,right delimiter={]}, matrix of math nodes, font=\footnotesize, column sep=0mm,nodes={inner sep=0.5mm}] 
         { 0 & 0 & 0 & 0 & 1 & 0 & 0 & 1 & 0 & 0 & 0 & 0 \\ };
         
\matrix [left delimiter=[,right delimiter={]}, matrix of math nodes, outer sep=0pt, font=\footnotesize, column sep=0mm,nodes={inner sep=0.5mm}]  at (\Xdroit,\Ydroit)
         {
         0 & 0 & 0 & 0 & 0 & 1 & 0 & 0 & 0 & 0 & 0 & 0 \\
         0 & 0 & 0 & 0 & 1 & 0 & 0 & 0 & 0 & 0 & 0 & 0 \\
         0 & 0 & 0 & 0 & 0 & 0 & 0 & 0 & 0 & 0 & 0 & 1 \\
         0 & 0 & 1 & 0 & 0 & 0 & 0 & 0 & 0 & 0 & 0 & 0 \\
         0 & 0 & 0 & 0 & 0 & 0 & 0 & 0 & 0 & 1 & 0 & 0 \\
         0 & 0 & 0 & 0 & 0 & 0 & 0 & 0 & 1 & 0 & 0 & 0 \\
         1 & 0 & 0 & 0 & 0 & 0 & 0 & 0 & 0 & 0 & 0 & 0 \\
         0 & 0 & 0 & 1 & 0 & 0 & 0 & 0 & 0 & 0 & 0 & 0 \\
         0 & 0 & 0 & 0 & 0 & 0 & 1 & 0 & 0 & 0 & 0 & 0 \\
         0 & 0 & 0 & 0 & 0 & 0 & 0 & 0 & 0 & 0 & 1 & 0 \\
         0 & 0 & 0 & 0 & 0 & 0 & 0 & 1 & 0 & 0 & 0 & 0 \\
         0 & 1 & 0 & 0 & 0 & 0 & 0 & 0 & 0 & 0 & 0 & 0 \\
	      }; at (10,10);

\matrix (m) [left delimiter=[,right delimiter={]}, matrix of math nodes, font=\footnotesize, column sep=0mm,nodes={inner sep=0.5mm}] at (\Xdroit,0)
         { 0 & 0 & 0 & 1 & 0 & 0 & 0 & 0 & 0 & 1 & 0 & 0 \\ };
         
\node at (1.65,0.63) {×};

\begin{pgfonlayer}{background}
\draw[line width=2mm, draw=red!60, line cap=round] ({\Xdroit * (-0.10)},{\Ydroit * (-0.12)}) -- ({\Xdroit * (-0.10)},{\Ydroit * 1.18}) -- ({\Xdroit * 1.5},{\Ydroit * 1.18});
\draw[line width=2mm, fill=red!60, red!60] ({\Xdroit * (1.232)},0.15) rectangle ++(0,-0.30);

\draw[line width=2mm, draw=green!25, line cap=round] ({\Xdroit * (0.10)},{\Ydroit * (-0.12)}) -- ({\Xdroit * (0.10)},{\Ydroit * 0.82}) -- ({\Xdroit * 1.5},{\Ydroit * 0.82});
\draw[line width=2mm, fill=green!25, green!25] ({\Xdroit * (0.833)},0.15) rectangle ++(0,-0.30);
\end{pgfonlayer}
\end{tikzpicture}

\end{center}
\caption{\small
  Illustration of multiplication between a vector and a permutation matrix, here
  with $n=12$.
}
\label{fig:mul_permu}
\end{figure}

Figure~\ref{fig:mul_permu} illustrates the multiplication of a vector by a
permutation matrix, which should be considered. The conventional method of
implementation is to loop over each bit element of the vector and perform, if
the current bit is set, a XOR operation between the corresponding line of the
matrix, and an accumulator. It is important to use the XOR operation specifically
rather than addition, because we are working in the $\mathbb{F}_2$ finite field.
As previously mentioned, the Hamming weight of the accumulator vector at the end
of this process should be the same as that of the input. This way of
implementing matrix multiplications is common and is described
in~\cite{SPA_McEliece}. We are targeting this implementation with the scope of
fault injection attacks.

As a side note, since the McEliece cryptosystem involves keys with
significant sizes, optimizing memory management is paramount. For
example, using the standard \texttt{bool} type, or a whole variable to
store each of the bits would be wasteful: on a 32-bit processor, it
would imply to ignore 31 out of the 32 bits of variables.

Instead, a preferred approach is to fill the whole variables with useful bits
only, even if it means using bitwise operations, and bit masking. This method is
more efficient, and becomes mandatory on a target platform with memory
limitations. This is often the case on 8-bit systems, which have been targeted
by some implementations~\cite{microeliece}. While these constraints are not as
hard on more powerful or usual platforms, such as computers based on a x86
processor, wasting 31 bits out of 32 is still quite inefficient. In this
article, the size of the variables in bits will be denoted as $p$.

\begin{figure}[hbt]
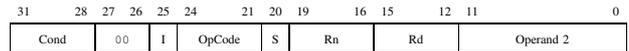

\resizebox{\linewidth}{!}{
  \renewcommand{\arraystretch}{2}
  \begin{tabular}{lllllllllllllllllllllllllllllll}
   31 & & & 28 & 27 & 26 & 25 & 24 & & & 21 & 20 & 19 & & & 16 & 15 & & & 12 & 11 & & & & & & & & & 0\\
    \hline
  \multicolumn{4}{| c |}{Cond} &
  \multicolumn{2}{| c |}{\texttt{00}} &
  \multicolumn{1}{| c |}{I} &
  \multicolumn{4}{| c |}{OpCode} &
  \multicolumn{1}{| c |}{S} &
  \multicolumn{4}{| c |}{Rn} &
  \multicolumn{4}{| c |}{Rd} &
  \multicolumn{10}{| c |}{Operand 2}\\
    \hline
  \end{tabular}
}
\caption{\small
  Representation of a 32-bits instruction designed for ARM
  processors~\cite{armv8manual}.
}
\label{fig:ARM_instru}
\end{figure}

In this attack, we target the XOR operation between the accumulator and specific
lines of the matrix. In the ARM instruction set~\cite{armv8manual}, instruction
representation contains a 4-bit long operation codes located between the 21st
and 24$^\text{th}$ bits, as shown in Figure~\ref{fig:ARM_instru}.

The operation code for XOR operation, called \texttt{EOR}, has a value
\texttt{0001}. By making a single bit change, we can switch it to an
\texttt{RSB} instruction with the value \texttt{0011}, which performs
subtraction between the operands. The use of the XOR operation is responsible
for preserving the Hamming weight property, as there is only one bit set in each
row of the matrix. For each one in the input vector, one and only one bit is set
in the result. Replacing it with subtraction eliminates this
property. The XOR operation is performed between the
accumulator and specific lines of the
matrix. A naive implementation is shown in Listing~\ref{lst:code_c}, with
$p = 32$, and $n = 1024$. For each bit in the input vector, we check if it is set, and if so, we
XOR the corresponding line of the matrix with the accumulator variable by
variable.

\begin{lstlisting}[language=C, caption={Naive vector-matrix multiplication in C-code.}, label={lst:code_c}]
uint32_t accu[1024/32] = {0};
for(int i = 0; i < 1024; i++) {
  if(((vector[i/32] >> (31-(i%32))) & 0x01) != 0) {
    for(int j = 0; j < (1024/32); j++) {
      accu[j] = accu[j] ^ matrix[i*(1024/32)+j];
}}}
\end{lstlisting}

The ARM built version of Listing~\ref{lst:code_c} is shown in
Listing~\ref{lst:code_asm}. In this listing, the instruction of interest is
located at address \texttt{0x10698} (in red in Listing~\ref{lst:code_asm}).

\begin{lstlisting}[%
 basicstyle=\small,
 label={lst:code_asm},
 caption={Built version of vulnerable code from Listing~\ref{lst:code_c} into ARM-assembly.},
 language={[ARM]Assembler},
 basicstyle = \footnotesize\ttfamily,
  escapechar=\%,
]
@ |               Instruction                  |
@ | Add. | bin. value | mnemonic               |
   10660   e51b300c     ldr r3, [fp, #-12]
   10664   e1a03103     lsl r3, r3, #2
   10668   e24b2004     sub r2, fp, #4
   1066c   e0823003     add r3, r2, r3
   10670   e5131024     ldr r1, [r3, #-36]
   10674   e51b2008     ldr r2, [fp, #-8]
   10678   e1a03002     mov r3, r2
   1067c   e1a03083     lsl r3, r3, #1
   10680   e0832002     add r2, r3, r2
   10684   e51b300c     ldr r3, [fp, #-12]
   10688   e0822003     add r2, r2, r3
   1068c   e59f30d8     ldr r3, [pc, #216]
   10690   e08f3003     add r3, pc, r3
   10694   e7933102     ldr r3, [r3, r2, %lsl% #2]
   %\highlightcode{red}{10698}%   %\highlightcode{red}{e0212003}%     %\highlightcode{red}{\textbf{eor} r2, r1, r3}%
   1069c   e51b300c     ldr r3, [fp, #-12]
   106a0   e1a03103     lsl r3, r3, #2
   106a4   e24b1004     sub r1, fp, #4
   106a8   e0813003     add r3, r1, r3
   106ac   e5032024     str r2, [r3, #-36]
\end{lstlisting}

By changing only one bit, we can shift from \texttt{EOR} instruction
(\texttt{0xe0\textbf{2}12003}) to \texttt{RSB} instruction
(\texttt{0xe0\textbf{6}12003}), whose provides a subtraction between the
operands.\newline

Thus, our goal is to cause corruption at the end of the global output, depending
on attributes of the multiplication result. The considered fault
model implies a one-bit change in the program's instructions, either before they
are loaded in volatile memory (in RAM), or in the processor caches, because we
have both temporal and spatial proximity: the selected instruction is likely to
be in a \texttt{for} loop executed many times, with few other manipulations
inside.

An important property of linear correcting codes is that a bit vector filled
with zeros is inevitably a valid code. Because the coding operation can be
performed using only matrix multiplication, it is necessary the case that a null
vector is the only code that can possibly represents a null vector.

During normal operations, sending input vectors with a Hamming weight of less
than $t$ should not provoke any error or message corruption. With the modified
instruction, this is completely possible. We thus propose the following approach:

\begin{enumerate}
  \item send an input vector with an Hamming weight of
        $\left\lceil \dfrac{t}{p} \right\rceil$;
  \item observe the occurrence or absence of corruption or error.
\end{enumerate}

In step 1, the positions of the set bits can be randomly selected. In step 2,
the algorithm's failure to produce correct results indicates that the set bits
in the rows of the permutation matrix corresponding to the input vector are not
grouped together in the same $p$ columns. Specifically, if the set bits were in
the same group, it would be impossible to end up with an accumulator that has a
Hamming weight above $t$.

This approach provides crucial data about the permutation matrix. By repeating
it many times with different inputs, one can considerably reduce the size of the
possible matrices, and each of the remaining matrices can be tested by
decrypting valid vectors with a high Hamming weight and looking for errors. If
one knows the permutation matrix, it becomes possible to attack the
error-correcting and scrambling matrices, as described in~\cite{SPA_McEliece},
enabling complete recovery of the private key.

\subsection{Details on a single iteration}
In order to ease understanding, let's focus on an example multiplication with
small parameters: $n = 12$, $p = 4$, and $t = 5$. Since we have $\lceil
\frac{5}{4} \rceil$, we need to input vectors with a Hamming weight of 2. This
is shown in Figure~\ref{fig:mul_permu}, which illustrates an unmodified
process. As expected, the output vector has the same Hamming weight as the
input.

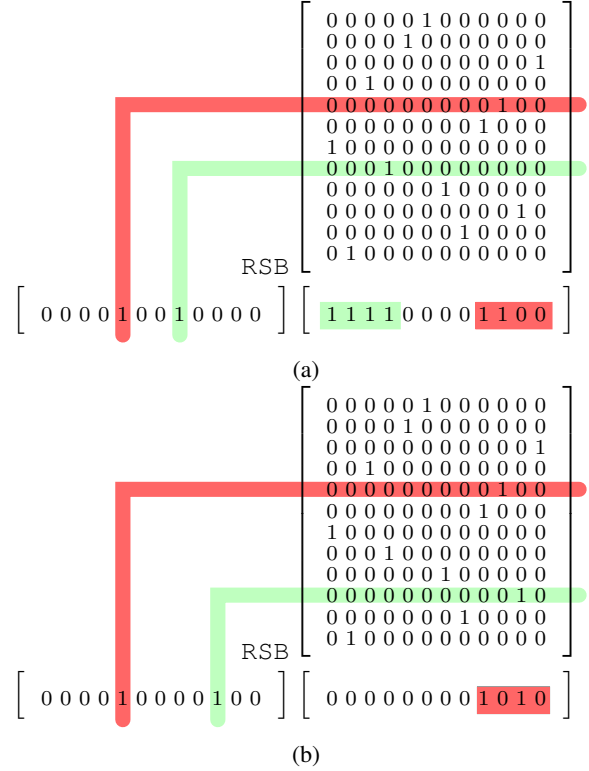
\begin{figure}[hbt]
  \begin{center}
    \begin{subfigure}{0.45\textwidth}
      \begin{tikzpicture}
\def\Xdroit{3.75}
\def\Ydroit{2.35}

\matrix (m) [left delimiter=[,right delimiter={]}, matrix of math nodes, font=\footnotesize, column sep=0mm,nodes={inner sep=0.5mm}] 
         { 0 & 0 & 0 & 0 & 1 & 0 & 0 & 1 & 0 & 0 & 0 & 0 \\ };
         
\matrix [left delimiter=[,right delimiter={]}, matrix of math nodes, outer sep=0pt, font=\footnotesize, column sep=0mm,nodes={inner sep=0.5mm}]  at (\Xdroit,\Ydroit)
         {
         0 & 0 & 0 & 0 & 0 & 1 & 0 & 0 & 0 & 0 & 0 & 0 \\
         0 & 0 & 0 & 0 & 1 & 0 & 0 & 0 & 0 & 0 & 0 & 0 \\
         0 & 0 & 0 & 0 & 0 & 0 & 0 & 0 & 0 & 0 & 0 & 1 \\
         0 & 0 & 1 & 0 & 0 & 0 & 0 & 0 & 0 & 0 & 0 & 0 \\
         0 & 0 & 0 & 0 & 0 & 0 & 0 & 0 & 0 & 1 & 0 & 0 \\
         0 & 0 & 0 & 0 & 0 & 0 & 0 & 0 & 1 & 0 & 0 & 0 \\
         1 & 0 & 0 & 0 & 0 & 0 & 0 & 0 & 0 & 0 & 0 & 0 \\
         0 & 0 & 0 & 1 & 0 & 0 & 0 & 0 & 0 & 0 & 0 & 0 \\
         0 & 0 & 0 & 0 & 0 & 0 & 1 & 0 & 0 & 0 & 0 & 0 \\
         0 & 0 & 0 & 0 & 0 & 0 & 0 & 0 & 0 & 0 & 1 & 0 \\
         0 & 0 & 0 & 0 & 0 & 0 & 0 & 1 & 0 & 0 & 0 & 0 \\
         0 & 1 & 0 & 0 & 0 & 0 & 0 & 0 & 0 & 0 & 0 & 0 \\
	      }; at (10,10);

\matrix (m) [left delimiter=[,right delimiter={]}, matrix of math nodes, font=\footnotesize, column sep=0mm,nodes={inner sep=0.5mm}] at (\Xdroit,0)
         { 1 & 1 & 1 & 1 & 0 & 0 & 0 & 0 & 1 & 1 & 0 & 0 \\ };
         
\node at (1.5,0.63) {\texttt{RSB}};

\begin{pgfonlayer}{background}
\draw[line width=2mm, draw=red!60, line cap=round] ({\Xdroit * (-0.10)},{\Ydroit * (-0.12)}) -- ({\Xdroit * (-0.10)},{\Ydroit * 1.18}) -- ({\Xdroit * 1.5},{\Ydroit * 1.18});
\draw[red!60, fill=red!60] ({\Xdroit * (1.165)-0.10}, 0.15) rectangle ++(1.00,-0.35);

\draw[line width=2mm, draw=green!25, line cap=round] ({\Xdroit * (0.10)},{\Ydroit * (-0.12)}) -- ({\Xdroit * (0.10)},{\Ydroit * 0.82}) -- ({\Xdroit * 1.5},{\Ydroit * 0.82});
\draw[green!25, fill=green!25] ({\Xdroit * (0.833)-0.90}, 0.15) rectangle ++(1.05,-0.35);
\end{pgfonlayer}

\end{tikzpicture}

      \caption{}
      \label{fig:rsb_permu_err}
    \end{subfigure}
    \hfill
    \begin{subfigure}{0.45\textwidth}
      \begin{tikzpicture}[
  Deco/.style={postaction={
      pattern={Lines[angle=-45, distance=0.75mm,  line width=1mm]},
      pattern color=red!60,
    }},
  ]

\def\Xdroit{3.75}
\def\Ydroit{2.35}

\matrix (m) [left delimiter=[,right delimiter={]}, matrix of math nodes, font=\footnotesize, column sep=0mm,nodes={inner sep=0.5mm}] 
         { 0 & 0 & 0 & 0 & 1 & 0 & 0 & 0 & 0 & 1 & 0 & 0 \\ };
         
\matrix [left delimiter=[,right delimiter={]}, matrix of math nodes, outer sep=0pt, font=\footnotesize, column sep=0mm,nodes={inner sep=0.5mm}]  at (\Xdroit,\Ydroit)
         {
         0 & 0 & 0 & 0 & 0 & 1 & 0 & 0 & 0 & 0 & 0 & 0 \\
         0 & 0 & 0 & 0 & 1 & 0 & 0 & 0 & 0 & 0 & 0 & 0 \\
         0 & 0 & 0 & 0 & 0 & 0 & 0 & 0 & 0 & 0 & 0 & 1 \\
         0 & 0 & 1 & 0 & 0 & 0 & 0 & 0 & 0 & 0 & 0 & 0 \\
         0 & 0 & 0 & 0 & 0 & 0 & 0 & 0 & 0 & 1 & 0 & 0 \\
         0 & 0 & 0 & 0 & 0 & 0 & 0 & 0 & 1 & 0 & 0 & 0 \\
         1 & 0 & 0 & 0 & 0 & 0 & 0 & 0 & 0 & 0 & 0 & 0 \\
         0 & 0 & 0 & 1 & 0 & 0 & 0 & 0 & 0 & 0 & 0 & 0 \\
         0 & 0 & 0 & 0 & 0 & 0 & 1 & 0 & 0 & 0 & 0 & 0 \\
         0 & 0 & 0 & 0 & 0 & 0 & 0 & 0 & 0 & 0 & 1 & 0 \\
         0 & 0 & 0 & 0 & 0 & 0 & 0 & 1 & 0 & 0 & 0 & 0 \\
         0 & 1 & 0 & 0 & 0 & 0 & 0 & 0 & 0 & 0 & 0 & 0 \\
	      }; at (10,10);

\matrix (m) [left delimiter=[,right delimiter={]}, matrix of math nodes, font=\footnotesize, column sep=0mm,nodes={inner sep=0.5mm}] at (\Xdroit,0)
         { 0 & 0 & 0 & 0 & 0 & 0 & 0 & 0 & 1 & 0 & 1 & 0 \\ };
         
\node at (1.5,0.63) {\texttt{RSB}};

\begin{pgfonlayer}{background}
\draw[line width=2mm, draw=red!60, line cap=round] ({\Xdroit * (-0.10)},{\Ydroit * (-0.12)}) -- ({\Xdroit * (-0.10)},{\Ydroit * 1.18}) -- ({\Xdroit * 1.5},{\Ydroit * 1.18});

\draw[line width=2mm, draw=green!25, line cap=round] ({\Xdroit * (0.233)},{\Ydroit * (-0.12)}) -- ({\Xdroit * (0.233)},{\Ydroit * 0.583}) -- ({\Xdroit * 1.5},{\Ydroit * 0.583});

\draw[Deco, fill=green!25] ({\Xdroit * (1.165)-0.08}, 0.15) rectangle ++({\Xdroit * (0.21)+0.16},-0.35);
\end{pgfonlayer}

\end{tikzpicture}

      \caption{}
      \label{fig:rsb_permu_ok}
    \end{subfigure}
  \end{center}
  \caption{\small
    Illustration of faulty multiplications between a vector and a
    permutation matrix, with \texttt{EOR} replaced by \texttt{RSB}.
    Figure~\ref{fig:rsb_permu_err} triggers an error or corruption, whereas
    Figure~\ref{fig:rsb_permu_ok} does not.
  }
  \label{fig:rsb_permu}
\end{figure}

Now let's consider a process that has encountered a fault, indicated by the
replacement of the XOR instruction with an \texttt{RSB}. This operation
is depicted in Figure~\ref{fig:rsb_permu}. For each row processed in the
matrix, 4 bits will be subtracted from the accumulator at a time.
The Hamming weight of the output vector has been modified. In
Figure~\ref{fig:rsb_permu_err}, the first bit set in the input vector now
generates 2 ones, because $0 - 4 = -4$, and the second one generates 4 ones,
because $0 - 1 = -1$. Note that in ARM, negative numbers are represented using
two's complement. Conceptually, we can also consider these as 4-bit unsigned
variable. Despite the integer overflow, the resulting bits are the same.
Additionally, carries are not applied between variables, so empty groups will
remain empty.

The Hamming weight of the output vector has increased to 6, which is above $t$,
and will result in an error or corruption at the end of the global algorithm.
This indicates that the matrix's lines corresponding to the ones in the input
vector have their set bits in different 4-bit columns. In
Figure~\ref{fig:rsb_permu_ok}, we illustrate a case where the set bits of the
relevant rows are in the same 4-bit group, resulting in an output vector with
only two set bits. The maximum Hamming weight we could theoretically have in the
output is $p$, since only one variable would be affected. In such a case, the
correcting code would be able to retrieve the null vector.

By reiterating this operation with various input vectors with the same weight,
one can group the matrix's rows going in the same group, without knowing which
ones. This represents a significant reduction in the space of the possible
permutation matrices.

\subsection{Resulting metrics}
\label{sec:resultat-exp}

By exploiting this attack, the key space can be significantly
reduced. The extent of reduction varies greatly depending on the
algorithm's parameter and the size of data on the target platform. As
for the permutation matrix, it has a size of $n \times n$ bits. Due to
its nature, the associated entropy is not $log_2(2^{n^2})$ as it would
be with a regular matrix, but $log_2(n!)$, which is the number of
possible permutations of the elements of an $n$-bit vector. In the
McEliece cryptosystem, the original specification provides example
parameters with $n= 1024$~\cite{mceliece_spec}, but today, bigger
sizes are commonly used. The PQCRYPTO recommendations mention
parameters with $n = 6960$~\cite[section 4]{pqcrypto}.

The above approach determines whether the selected rows each have their set bit
in different groups of $p$ columns. However, by repeatedly iterating it, sets of
rows having their set bit in the same group can be formed without being able to
identify which rows belong to which set. With enough iterations, obtaining a
complete partition becomes achievable. In this case, the remaining entropy is
$log_2(p!^{\frac{n}{p}} \times \frac{n}{p}!)$ when $\frac{n}{p}$ is an integer.

\begin{figure}[hbt]
  \begin{center}
    \begin{tikzpicture}
\begin{axis}[
	symbolic x coords={8,16,32,64,$\emptyset$},
	ytick distance=1000,
    ymajorgrids=true,
    ymin = 0,
    ymax = 9000,
    bar width=0.6cm,
    height=6.25cm,
    width=7.5cm,
    xlabel = {$p$},
    ylabel = {Entropy in bits}]
\addplot[ybar,fill=olive]
	coordinates {(8,2674.460348) (16,3128.004134)
		 (32,3882.887701) (64,4780.172444) ($\emptyset$,8769.006144)};
\end{axis}
\end{tikzpicture}
  \end{center}
  \caption{\small
    Comparison of the permutation matrix's entropy before and after the
    attack, with multiple $p$ values, for $n = 1024$. The $\emptyset$ value
    occurred prior to the attack.
  }
  \label{fig:entropie_1}
\end{figure}
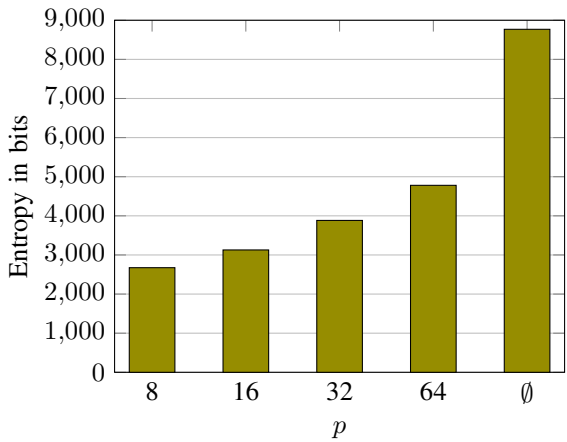

As expected, the entropy decreases with smaller variables size.
Figure~\ref{fig:entropie_1} shows entropy values for the parameter $n = 1024$,
which is the one suggested in the original McEliece specification, but for
different variables size. The $\emptyset$ value represents the entropy before
the attack. Even with 64-bit registers, a significant reduction in possibilities
occurs.

\begin{figure}[!hbt]
  \begin{center}
    \pgfplotsset{scaled y ticks=false}
\begin{tikzpicture}
\begin{axis}[
	symbolic x coords={8,16,32,64,$\emptyset$},
	ytick distance=10000,
    ymajorgrids=true,
    ymin = 0,
    ymax = 80000,
    bar width=0.6cm,
    height=6.25cm, width=7.5cm,
    xlabel = {$p$},
    ylabel = {Entropy in bits}]
\addplot[ybar,fill=olive]
	coordinates {(8,20556.8129) (16,22439.66636)
		 (32,26972.04198) (64,32772.96789) ($\emptyset$,78810.05699)};
\end{axis}
\end{tikzpicture}
  \end{center}
  \caption{\small
    Comparison of the permutation matrix's entropy before and after the
    attack, with multiple $p$ values, for $n = 6960$. The $\emptyset$ value
    occurred prior to the attack.
  }
  \label{fig:entropie_2}
\end{figure}
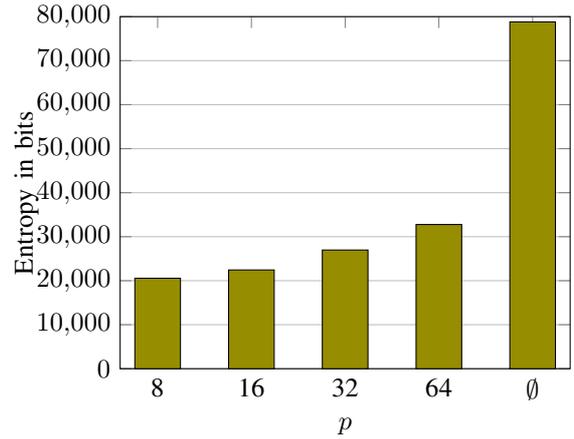

The relative proportions of remaining entropy remain in the same order of
magnitude even when $n$ varies, as observed in Figure~\ref{fig:entropie_2},
which shows the results for $n = 6960$, the suggested size in the PQCRYPTO
recommendations~\cite[section 4]{pqcrypto}.

In our experiments, only the \texttt{RSB} instruction has been
considered as it is the only operation code reachable with only one
bit change having desirable effects. Indeed, \texttt{AND} and
\texttt{TEQ} cannot induce any Hamming weight change in the
accumulator variable. Such an injection has been demonstrated as
possible in hardware~\cite{cayrel_HDR}. It is also reproducible in
software using binary instrumentation tools. In the latter case, there
is no reason to restrict the change to one bit. Exploring all the
other operation codes can possibly lead to increased performances, by
developing and exploiting a different routine. In hardware attacks,
affecting multiple bits is plausible, and expecting bigger changes in
the operation code can be conceivable.

The transposition of this hardware-based attack to the software context is not
only feasible but raises many questions, particularly on its impact over
white-box implementations.

\section{Exploitation of the fault attack on implementation of the McEliece cryptosystem}
\label{sec:discussion}

We introduced the first fault injection based attack on the McEliece cryptosystem.
This presented attack in the previous sections can be applied since an attacker
is able to shift a bit, on the one hand, on the typical implementation on a
component through fault injection attack~\cite{journals/access/BreierH22} or, on
the other hand, with binary instrumentation on software application on an
uncontrolled environment. Our attack does not target the McEliece cryptosystem
specification, but rather a conventional method of implementing
it~\cite{SPA_McEliece}, specifically the depermutation of the input.

When one is in possession of an unobfuscated executable applying the
decryption, recovering the key is trivial. Hardware attacks are
necessary when it is embedded on a platform on which we cannot run
debuggers, for example. With the advent of the white-box security
model, however, many industrial actors released obfuscated executables
with anti-instrumentation measures, with the hope that an attacker
with binary exploitation tools could not recover secrets. If these are
not advanced enough, the attack presented here can circumvent them.

Implementation of cryptographic algorithms following the \ac{WBC} security model
are usually based on internal encoding applied to precomputed tables. In such
cases, an attack based on data manipulation is called into question: \textit{do the
modifications still hold through random bijections}? When the attack is based on
instructions manipulation, it seems jeopardized when considering implementations
based on precomputation. The targeted instructions are not in the executable,
but are exploited in advance during generation of the software. Therefore, it
needs to be converted to an attack based on data.

When considering a table of precomputed data, re-computing it can be
envisioned. Indeed, the results are the content itself, and one of the two
operands of the associated operation is the corresponding index of each value.
Using the same example of the XOR instruction, one can simply apply an
exclusive-or between the result and its index to recover the second parameter,
and eventually recompute the whole table in a faulty way as explained if
necessary. This technique is applicable, notably at the end of a \ac{WBC}
protected algorithm without external encoding. It should be noted that it is
possible when an implementation follows an open specification: the knowledge
required is not only the result and one operand, but also the operation applied
itself.

An attack such as the one presented in this study is very likely to affect
naively obfuscated implementations without precomputed intermediary results.
Control-flow obfuscation and the addition of useless calculation
around the desired one do not prevent the existence of the XOR
instruction, which is our target of choice.

These implications have major consequences in the exploitation of embedded
wireless network capabilities. All industrial actors have interests in providing
solutions with a low use of the mobile network usage, as it is expensive for the
end-user and its availability cannot be assumed. The existence of white spots
makes it hard to rely too much on such communications. The white-box
implementation of an algorithm is thus expected to provide many features, such
as low data size to reduce the burden of its downloads and robustness, since an
easily attackable executable will need to be changed more often, and this will
imply more downloads too. These features are all related to the stakes around
telecommunications use, and they draw a direct, consequential link between
security and mobile networks constraints.

\section{A variant of the McEliece cryptosystem immuned to our attack}
\label{sec:wbc-pq}

Our work focuses on decryption in the McEliece specification. Since encryption
relies on a public key, attempting to protect it with a \ac{WBC} security model
is unproductive. Therefore, we aim at protecting the private key during the
deciphering process. As illustrated in Fi\-gu\-re~\ref{fig:McEliece_internals}, this
involves hiding the permutation matrix $P$, the scrambling matrix $S$, and the
generator matrix $G$ of the underlying linear code.

One might be tempted to simply merge the permutation and scrambling matrices
each with random linear applications that could be cancelled outside of the
algorithm. However, this method introduces risks towards attacks based on
interpolation. To properly protect these assets, the techniques used cannot rely
exclusively on linear fusions.

Linear mathematical operations, which make up almost all of the McEliece
cryptosystem, are easily precomputable, despite the potentially large size of
the associated matrices. This was done, for example, in the original proposition
for a white-box version of the \ac{AES}~\cite{WB_AES} when considering the
\texttt{MixColumns} operation.

The usual technique involves decomposing the entire matrix into smaller ones, as
shown in Figure~\ref{fig:decompo_matrice}, with a matrix named $M$, where $a$ is
the number of submatrices rows, and $b$ is the number of submatrices columns.
Their corresponding linear applications are then precomputed. The size of the
submatrices is completely up to the designer and will influence usability:
larger submatrices will result in larger precomputed tables and thus a larger
white-box implementation. Choosing a submatrix size that is not a total matrix
size divider is possible.

\begin{figure}[hbt]
  \begin{center}
    \begin{tikzpicture}
\def\matAX{-2.25}
\def\matAY{2.07}
\def\matBX{-1.05}
\def\matBY{2.07}
\def\matCX{-2.25}
\def\matCY{1}
\def\matDX{2.25}
\def\matDY{2.07}
\def\matEX{-2.25}
\def\matEY{-2.1}
\def\matFX{2.25}
\def\matFY{-2.1}
\def\matLargeur{0.5}
\def\matHauteur{3.25}

\matrix [left delimiter=[,right delimiter={]}, matrix of math nodes, outer sep=0pt, column sep=1mm, row sep=3.5mm, nodes={inner sep=0.5mm}, font=\huge] at (0,0)
         {
         ~ & ~ & ~ & ~ & ~ & ~ & ~ & ~ & ~ & ~ & ~ & ~ \\
         ~ & ~ & ~ & ~ & ~ & ~ & ~ & ~ & ~ & ~ & ~ & ~ \\
         ~ & ~ & ~ & ~ & ~ & ~ & ~ & ~ & ~ & ~ & ~ & ~ \\
         ~ & ~ & ~ & ~ & ~ & ~ & ~ & ~ & ~ & ~ & ~ & ~ \\
         ~ & ~ & ~ & ~ & ~ & ~ & ~ & ~ & ~ & ~ & ~ & ~ \\
         ~ & ~ & ~ & ~ & ~ & ~ & ~ & ~ & ~ & ~ & ~ & ~ \\
         ~ & ~ & ~ & ~ & ~ & ~ & ~ & ~ & ~ & ~ & ~ & ~ \\
         ~ & ~ & ~ & ~ & ~ & ~ & ~ & ~ & ~ & ~ & ~ & ~ \\
         ~ & ~ & ~ & ~ & ~ & ~ & ~ & ~ & ~ & ~ & ~ & ~ \\
         ~ & ~ & ~ & ~ & ~ & ~ & ~ & ~ & ~ & ~ & ~ & ~ \\
         ~ & ~ & ~ & ~ & ~ & ~ & ~ & ~ & ~ & ~ & ~ & ~ \\
         ~ & ~ & ~ & ~ & ~ & ~ & ~ & ~ & ~ & ~ & ~ & ~ \\
	      }; 

\node at (\matAX - \matLargeur, \matAY) {\scalebox{1}[\matHauteur]{[}};
\node at (\matAX + \matLargeur, \matAY) {\scalebox{1}[\matHauteur]{]}};
\node at (\matAX, \matAY) {$M_{0,0}$};

\node at (\matBX - \matLargeur, \matBY) {\scalebox{1}[\matHauteur]{[}};
\node at (\matBX + \matLargeur, \matBY) {\scalebox{1}[\matHauteur]{]}};
\node at (\matBX, \matBY) {$M_{0,1}$};

\node at (\matCX - \matLargeur, \matCY) {\scalebox{1}[\matHauteur]{[}};
\node at (\matCX + \matLargeur, \matCY) {\scalebox{1}[\matHauteur]{]}};
\node at (\matCX, \matCY) {$M_{1,0}$};

\node at (\matDX - \matLargeur, \matDY) {\scalebox{1}[\matHauteur]{[}};
\node at (\matDX + \matLargeur, \matDY) {\scalebox{1}[\matHauteur]{]}};
\node at (\matDX, \matDY) {$M_{0,b}$};

\node at (\matEX - \matLargeur, \matEY) {\scalebox{1}[\matHauteur]{[}};
\node at (\matEX + \matLargeur, \matEY) {\scalebox{1}[\matHauteur]{]}};
\node at (\matEX, \matEY) {$M_{a,0}$};

\node at (\matFX - \matLargeur, \matFY) {\scalebox{1}[\matHauteur]{[}};
\node at (\matFX + \matLargeur, \matFY) {\scalebox{1}[\matHauteur]{]}};
\node at (\matFX, \matFY) {$M_{a,b}$};

\node at (-3.5,0) {$M =$};

\draw[dotted] (\matCX,\matCY - 1) -- (\matEX,\matEY + 1);
\draw[dotted] (\matCX + 1,\matBY - 1) -- (\matFX - 1,\matFY + 1);
\draw[dotted] (\matBX + 1,\matBY) -- (\matDX - 1,\matDY);

\end{tikzpicture}
  \end{center}
  \caption{\small
    Visualization illustrating the decomposition of a matrix in order to
    make the precomputation of its associated transformation possible.
  }
  \label{fig:decompo_matrice}
\end{figure}

The resulting tables can be used consecutively by adding their results. If we
consider a vector $u$ multiplied with $M$ on its left, then resulting $u \times M$ is
composed of $b$ concatenated subvectors:

\begin{align*}
u \times M = \left[ \sum_{i=0}^{a} u_i \times M_{i,0} ~ \| ~ \sum_{i=0}^{a} u_i \times M_{i,1} ~ \| ~ ... \right. \\ \left. ... ~ \| ~ \sum_{i=0}^{a} u_i \times M_{i,b} \right]
\end{align*}

The addition themselves should also be precomputed, so that the entire linear
application consists solely of tables. Each subvector thus becomes the root
result of a precomputed tree. In the context of the McEliece cryptosystem, the
XOR operations are used for additions. A tables tree of this kind can be
deployed to handle the permutation and scrambling matrix.

However, precomputation is not possible on decoding step in the decryption
process, since it operates on an input that is often at least 1024 bits long. As
a result, adaptation work is necessary to implement the McEliece cryptosystem in
a \ac{WBC} security model.

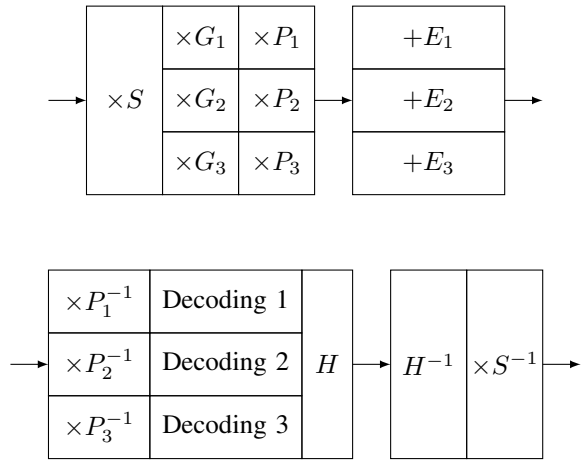
\begin{figure}[hbt]
\begin{center} \begin{tikzpicture}
\def\hauteurB{2.5}
\def\largeurB{2}
\def\largeurF{0.5}
\def\espaceV{3.5}
\def\decalageH{(\largeurB * (\coefDechif - \coefChif))/2}
\def\coefChif{1.5}
\def\coefDechif{2}

\draw({\largeurF + \decalageH}, \espaceV) rectangle ++({\largeurB * \coefChif * (1/3)}, \hauteurB) node[pos=0.5] {$\times S$}; 
\draw({\largeurF + \decalageH + \largeurB * \coefChif * (1/3)}, {\espaceV + (0/3)*\hauteurB}) rectangle ++({\largeurB * \coefChif * (1/3)}, {\hauteurB * (1/3)}) node[pos=0.5] {$\times G_3$};
\draw({\largeurF + \decalageH + \largeurB * \coefChif * (1/3)}, {\espaceV + (1/3)*\hauteurB}) rectangle ++({\largeurB * \coefChif * (1/3)}, {\hauteurB * (1/3)}) node[pos=0.5] {$\times G_2$};
\draw({\largeurF + \decalageH + \largeurB * \coefChif * (1/3)}, {\espaceV + (2/3)*\hauteurB}) rectangle ++({\largeurB * \coefChif * (1/3)}, {\hauteurB * (1/3)}) node[pos=0.5] {$\times G_1$};
\draw({\largeurF + \decalageH + \largeurB * \coefChif * (2/3)}, {\espaceV + (0/3)*\hauteurB}) rectangle ++({\largeurB * \coefChif * (1/3)}, {\hauteurB * (1/3)}) node[pos=0.5] {$\times P_3$};
\draw({\largeurF + \decalageH + \largeurB * \coefChif * (2/3)}, {\espaceV + (1/3)*\hauteurB}) rectangle ++({\largeurB * \coefChif * (1/3)}, {\hauteurB * (1/3)}) node[pos=0.5] {$\times P_2$};
\draw({\largeurF + \decalageH + \largeurB * \coefChif * (2/3)}, {\espaceV + (2/3)*\hauteurB}) rectangle ++({\largeurB * \coefChif * (1/3)}, {\hauteurB * (1/3)}) node[pos=0.5] {$\times P_1$};
\draw({\largeurB * \coefChif + \largeurF * 2 + \decalageH}, {\espaceV + (0/3)*\hauteurB}) rectangle ++({\largeurB}, {\hauteurB * (1/3)}) node[pos=0.5] {$+ E_3$};
\draw({\largeurB * \coefChif + \largeurF * 2 + \decalageH}, {\espaceV + (1/3)*\hauteurB}) rectangle ++({\largeurB}, {\hauteurB * (1/3)}) node[pos=0.5] {$+ E_2$};
\draw({\largeurB * \coefChif + \largeurF * 2 + \decalageH}, {\espaceV + (2/3)*\hauteurB}) rectangle ++({\largeurB}, {\hauteurB * (1/3)}) node[pos=0.5] {$+ E_1$};

\draw [-latex] ({\decalageH },{\hauteurB * 0.5 + \espaceV}) -- ++(\largeurF,0);
\draw [-latex] ({\largeurB * \coefChif + \largeurF * 1 + \decalageH},{\hauteurB * 0.5 + \espaceV}) -- ++(\largeurF,0);
\draw [-latex] ({\largeurB * (1 + \coefChif) + \largeurF * 2 + \decalageH},{\hauteurB * 0.5 + \espaceV}) -- ++(\largeurF,0);

\draw ({\largeurB * 0 + \largeurF * 1},{\hauteurB * (0/3)}) rectangle ++({\largeurB * \coefDechif * (1/3)},{\hauteurB * (1/3)}) node[pos=0.5] {$\times P_3^{-1}$};
\draw ({\largeurB * 0 + \largeurF * 1},{\hauteurB * (1/3)}) rectangle ++({\largeurB * \coefDechif * (1/3)},{\hauteurB * (1/3)}) node[pos=0.5] {$\times P_2^{-1}$};
\draw ({\largeurB * 0 + \largeurF * 1},{\hauteurB * (2/3)}) rectangle ++({\largeurB * \coefDechif * (1/3)},{\hauteurB * (1/3)}) node[pos=0.5] {$\times P_1^{-1}$};
\draw ({\largeurB * \coefDechif * (1/3) + \largeurF * 1},{\hauteurB * (0/3)}) rectangle ++({\largeurB * \coefDechif * (2/4)},{\hauteurB * (1/3)}) node[pos=0.5] {Decoding 3};
\draw ({\largeurB * \coefDechif * (1/3) + \largeurF * 1},{\hauteurB * (1/3)}) rectangle ++({\largeurB * \coefDechif * (2/4)},{\hauteurB * (1/3)}) node[pos=0.5] {Decoding 2};
\draw ({\largeurB * \coefDechif * (1/3) + \largeurF * 1},{\hauteurB * (2/3)}) rectangle ++({\largeurB * \coefDechif * (2/4)},{\hauteurB * (1/3)}) node[pos=0.5] {Decoding 1};
\draw ({\largeurB * \coefDechif * (5/6) + \largeurF * 1},0) rectangle ++({\largeurB * \coefDechif * (1/6)},\hauteurB) node[pos=0.5] {$H$};
\draw ({\largeurB * \coefDechif + \largeurF * 2},0) rectangle ++(\largeurB * 0.5,\hauteurB) node[pos=0.5] {$H^{-1}$};
\draw ({\largeurB * (\coefDechif + 0.5) + \largeurF * 2},0) rectangle ++(\largeurB * 0.5,\hauteurB) node[pos=0.5] {$\times S^{-1}$};

\draw [-latex] ({\largeurB * 0 + \largeurF * 0},\hauteurB * 0.5) -- ++(\largeurF,0);
\draw [-latex] ({\largeurB * \coefDechif + \largeurF * 1},\hauteurB * 0.5) -- ++(\largeurF,0);
\draw [-latex] ({\largeurB * (1 + \coefDechif ) + \largeurF * 2},\hauteurB * 0.5) -- ++(\largeurF,0);
\end{tikzpicture}
\end{center}
\caption{\small
  Representation of the operational principles of one explored
  McEliece cryptosystem modification. Above is the encryption, below is
  the decryption. In this example, the number of subcodes is set to 3.
}
\label{fig:McEliece_mod_internals}
\end{figure}

One possible approach, illustrated in
Figure~\ref{fig:McEliece_mod_internals}, is to use multiple small
correcting codes instead of a single large one. Each code would be
associated with its own permutation, but they would all share a common
scrambling matrix that would mix the data across the
subcodes. However, implementing this method would require
modifications not only on the decryption side but also to the
encryption process. On the later, the differences are almost invisible
to the final user: the subcodes are embedded in the public key, and
encrypting still essentially consists in multiplying by one large
matrix. However, adding errors must be applied considering each subcode.

To protect the private key during the deciphering step, internal
encoding based on random bijections, here named $H$, could be applied
at the end of the subcodes, with their inverses at the beginning of
the common unscrambling application. The subpermutations and subcodes
could be merged and precomputed together, while the scrambling matrix
would be subdivided and precomputed alone.

One crucial aspect to consider in this approach is that the permutation
matrices, which are critical to the security of the McEliece cryptosystem, may
have less entropy than the large equivalent one. This is due to the fast growth
of the factorial function that describes their complexity based on their size.
Therefore, the size of the permutation matrices must be chosen carefully, taking
into account both security and usability constraints. To ensure the
robustness of this idea, many offensive approaches still need to be
undertaken.

In this section, we introduce
a variant of McEliece cryptosystem aiming at being
protected against binary instrumentation, our variant should now be proved to be
resistant. This complex part is still an ongoing work.

\section{Conclusion}
\label{sec:conclusion}
In this article, we described the first fault injection based attack on ARM
implementation of the McEliece cryptosystem~\cite{SPA_McEliece} where we are
able to retrieve the private key. Our attack can be exploited anywhere where one
can modify a bit: through hardware attack on component or by binary instrumentation
on software executed in an uncontrolled environment. Our attack successfully
allows a substantial reduction of key space, regardless of the chosen
parameters.

We discussed how such an attack could be ported to afflict implementations
working in a \ac{WBC} security model. Given the numerous, powerful, and
accessible binary instrumentation tools available, we described how fault
injection and side channel attacks can be simulated and their consequences, to find new
vulnerability to mitigate.

We presented suggestions to mitigate the exploitability of our attack in an
uncontrolled environment. We also proposed ideas of variants on the McEliece
cryptosystem expected to lead to immune implementations so that they can protect
keys for a time long enough to allow practical use in real contexts, such as
embedded open devices fully mastered by malicious users. Network constraints and
limitations on binary updates sizes were major considerations in our design
process. However, precomputed tables and their associated sizes can still
represent a downside for effective use, notably due to bandwidth constraints.
Addressing this issue is crucial to find the most lightweight solution that can
effectively hold against binary instrumentation attacks.

Our proposed variant need further works to attest its viability in terms of
security. Also, more efforts could be dedicated into the estimation and the
management of this implementation's size: being able to make an informed
judgment about the trade-off between size, usability and security would be very
useful.

\section*{Acknowledgements}
The authors wish to thank David Naccache for his guidance and support all along
this work. They would also like to thank Aline Gouget and S\'ebastien Varrette
for their valuable comments.

\bibliographystyle{plain}
\bibliography{paper}

\balance{}


\begin{acronym}
  \acro{NIST}{National Institute of Standards and Technology}
  \acro{BYOD}{Bring Your Own Device}
  \acro{GSM}{Global System for Mobile Communications}
  \acro{WBC}{White-Box Cryptography}
  \acro{RSA}{Rivest–Shamir–Adleman}
  \acro{AES}{Advanced Encryption Standard}
  \acro{DPA}{Differential Power Analysis}
  \acro{DCA}{Differential Computational Analysis}
  \acro{DFA}{Differential Fault Analysis}
  \acro{COTS}{Customer Off-The-Shelf}
  \acro{GPU}{Graphics Processing Unit}
  \acro{SIMD}{Single Instruction on Multiple Data}
  \acro{BSI}{Bundesamt für Sicherheit in der Informationstechnik}
\end{acronym}

\end{document}